\documentclass[twocolumn,showpacs,preprintnumbers,amsmath,amssymb]{revtex4}

\usepackage{graphicx}
\usepackage{dcolumn}
\usepackage{bm}

\begin{document}


\title{Rearrangement of the vortex lattice due to instabilities of vortex flow}

\author{D.Y. Vodolazov$^{1,2}$}
\email{vodolazov@ipm.sci-nnov.ru}
\author{F.M. Peeters$^2$}
\email{francois.peeters@ua.ac.be} \affiliation{$^{1}$ Institute
for Physics of Microstructures, Russian Academy of Sciences,
603950, Nizhny Novgorod, GSP-105, Russia \\
$^2$Departement Fysica, Universiteit Antwerpen (CGB),
Groenenborgerlaan 171, B-2020 Antwerpen, Belgium}

\date{\today}


\begin{abstract}

With increasing applied current we show that the moving vortex
lattice changes its structure from a triangular one to a set of
parallel vortex rows in a pinning free superconductor. This effect
originates from the change of the shape of the vortex core due to
non-equilibrium effects (similar to the mechanism of vortex motion
instability in the Larkin-Ovchinnikov theory). The moving vortex
creates a deficit of quasiparticles in front of its motion and an
excess of quasiparticles behind the core of the moving vortex.
This results in the appearance of a wake (region with suppressed
order parameter) behind the vortex which attracts other vortices
resulting in an effective direction-dependent interaction between
vortices. When the vortex velocity $v$ reaches the critical value
$v_c$ quasi-phase slip lines (lines with fast vortex motion)
appear which may coexist with slowly moving vortices between such
lines. Our results are found within the framework of the
time-dependent Ginzburg-Landau equations and are strictly valid
when the coherence length $\xi(T)$ is larger or comparable with
the decay length $L_{in}$ of the non-equilibrium quasiparticle
distribution function. We qualitatively explain experiments on the
instability of vortex flow at low magnetic fields when the
distance between vortices $a \gg L_{in} \gg \xi (T)$. We speculate
that a similar instability of the vortex lattice should exist for
$v>v_c$ even when $a<L_{in}$.

\end{abstract}

\maketitle

\section{Introduction}

In 1976 Larkin and Ovchinnikov (LO) \cite{Larkin1} predicted an
instability of vortex motion that is related to the deviation of
the quasiparticle distribution function from its equilibrium value
near the vortex core (for review see \cite{Larkin2}). When the
vortex moves with a velocity $v$ the order parameter in the vortex
core varies on a time scale $\tau_{|\psi|} \sim \xi/v$ that can be
smaller than the relaxation time of the non-equilibrium
quasiparticles $\tau_{in}$ (due to inelastic electron-phonon or
electron-electron interactions). As a consequence, the
quasiparticle distribution function deviates from its equilibrium
and it results to a shrinkage of the vortex core at temperatures
close to the critical temperature \cite{Larkin2}. This effect is
mainly connected with the removal of quasiparticles from the
vortex core by the induced electric field \cite{Klein} and in some
respect is similar to the dynamic enhancement of superconductivity
in weak superconducting links \cite{Klein,Tinkham}.

Analytical calculations made in the 'dirty' limit (mean free path
length $\l$ of the electrons is smaller than the coherence length
$\xi$) predicted a decrease of the viscosity coefficient $\eta$ of
the vortex motion and the existence of a critical velocity $v_c
\sim 1/\sqrt{\tau_{in}}$ at which the viscous force $-\nu v$
reaches its maximal value \cite{Larkin2}. Macroscopically it
results into a nonlinear current-voltage characteristic $V \sim
I\eta(I)$ with pronounced hysteresis at relatively weak magnetic
fields \cite{Larkin2}. Such a behavior was experimentally observed
in many low \cite{Klein,Musienko,Lefloch,Babic} and high
\cite{Samoilov,Doettinger,Doettinger2,Xiao,Xiao2,Chiaverini}
temperature superconductors and quantitative agreement with theory
was found. From the experimental value of the critical voltage
$V_c=v_cBL$ ($B$ is a magnetic flux induction, L is the distance
between voltage leads) the critical velocity $v_c$ and relaxation
time $\tau_{in}$ were extracted.

We should stress the nontrivial nature of the LO effect. If we use
the Bardeen-Stephen expression \cite{Bardeen} for the viscosity of
the vortex motion $\eta=\Phi_0^2/2\pi \xi^2 \rho_n c$ we find that
it actually {\it increases} if the size of the vortex core
decreases (with for example a decrease of the temperature). In the
LO theory the vortex core shrinks and this results in a {\it
decrease} of the viscosity coefficient $\eta$. The possible
explanation of this contradiction is the failure of the
Bardeen-Stephen model for the vortex as a normal cylinder with
radius $\xi$ in case of a moving vortex with high enough velocity.

At low temperatures small changes in the quasiparticle
distribution function cannot influence the order parameter
\cite{Larkin2}. As a result nonlinear effects will start at larger
electrical fields and they become significant not only in the
vortex core but also around the vortex. Effectively, such
non-equilibrium effects were described as due to heating of the
quasiparticles by the induced electric field up to a temperature
larger than the sample/phonon temperature. This results in a
suppression of the order parameter near the vortex core and the
vortex core expands. This effect was used to explain the
experimental results on the vortex motion instability at low
temperatures for both 'dirty'
\cite{Babic,Kunchur,Kunchur2,Kunchur3} and 'pure' ($\l \gg \xi $)
superconductors \cite{Doettinger4}. In the latter case the
instability occurs due to a logarithmic dependence of the vortex
viscosity on the electronic temperature.

Returning to the LO theory we note that the main assumption here
was the uniform distribution of the non-equilibrium quasiparticle
distribution function $f(E)$ in the superconductor. (in particular
this leads to a field-independent critical velocity). From the
good {\it quantitative} agreement between theory and experiment in
large magnetic fields one may conclude that the above condition is
well satisfied when the distance between vortices satisfies $a(B)
\ll L_{in}$. However experiments at low magnetic fields showed
that this approach fails when $a(B)\sim \sqrt{\Phi_0/B}$ becomes
larger than $L_{in}$
\cite{Doettinger,Chiaverini,Doettinger3,Lefloch}. In Ref.
\cite{Doettinger3} it was proposed that the instability occurs
when the non-equilibrium distribution becomes uniform over the
superconductor which is possible if $v_c\tau_{in}\simeq a(B)$.
This leads to a dependence $v_c\sim \sqrt{1/B}$ at moderate
magnetic fields which was observed in many experiments
\cite{Doettinger,Chiaverini,Doettinger3,Lefloch}. But at lower
magnetic fields the critical velocity should behave as $v_c \sim
1/B$ to explain the field independent value for the critical
voltage $V_c$ \cite{Chiaverini} (the same conclusion can be drawn
from Fig. 10 of Ref. \cite{Klein}). Note that at these fields the
current induced magnetic field is still much less than the
external magnetic field and consequently it cannot explain the
observed effect.

Despite the large number of experimental works there is still the
fundamental question: what will happen with the vortex structure
when the critical velocity is approached and/or exceeded? In the
original paper of Larkin and Ovchinnikov it was assumed that the
vortex lattice does not exhibit any structural changes and
transits for $v>v_c$ to a state with a resistance close to the
normal value (in the current driven regime). However experiments
on low and high temperature superconductors showed that another
type of behavior is possible. For example, a transition to a state
with phase slip lines was experimentally observed in Ref.
\cite{Volotskaya,Volotskaya2,Dmitriev,Zybtsev,Sivakov} for low
temperature superconductors and similar IV characteristics (which
were differently interpreted) were observed in high temperature
superconductors in the voltage driven regime
\cite{Kunchur,Kunchur2,Kunchur3}. These experimental results
support the idea that some kind of phase transition occurs in the
vortex lattice at the instability point and regions with fast and
slow vortex motion appear in the sample
\cite{Volotskaya,Volotskaya2,Dmitriev,Zybtsev,Sivakov,Kunchur,Kunchur2,Kunchur3}.

To answer the above questions theoretically one should use a
rather complicated set of integro-differential equations for the
order parameter, Green functions of the superconductor and
quasiparticle distribution function \cite{Larkin2}. At the present
time it looks almost as an impossible task to solve these
equations even numerically. Therefore, we will limit ourselves to
the equations that were derived from the microscopic equations for
a superconductor in the dirty limit under the assumption that the
relaxation length $L_{in}=\sqrt{D\tau_{in}}$ of $f(E)$ ($D$ is the
diffusion constant) is smaller than the coherence length $\xi(T)$
\cite{Kramer,Watts-Tobin}. They are the extended (or generalized)
time-dependent Ginzburg-Landau equations and contain explicitly a
parameter $\tau_{in}$. From the very beginning we are in a
different limit as compared to the LO theory, because we consider
a non-equilibrium $f(E)$ that is nonuniform in the sample. The
longitudinal (odd in energy) part $f_L(E)=f(-E)-f(E)$ of the
non-equilibrium $f(E)$ distribution (which is actually responsible
for the variation of $|\psi|$) is localized only in the region
where the time derivative $\partial |\psi|/\partial t$ is finite
(see Eqs. (6,10) in Ref. \cite{Kramer}). It means that $f_L(E)$ is
non zero only near the moving vortex core.
\begin{figure}[hbtp]
\includegraphics[width=0.3\textwidth]{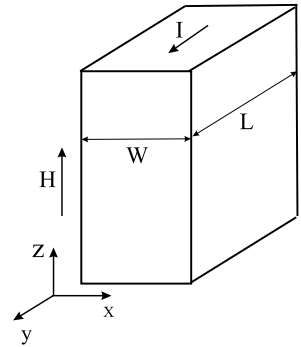}
\caption{Model system - a superconducting slab (infinitely long in
the y and z directions) in a parallel magnetic field $H$ with
transport current $I$.}
\end{figure}

The paper is organized as follows. In section II we present our
model system. In section III we study the rearrangement of the
vortex structure due to non-equilibrium effects at moderate and
high magnetic fields and in Sec. IV the non-equilibrium vortex
dynamics at zero magnetic field. Finally, in section V we discuss
our results and make a comparison with experiments.

\section{Model system}

As a model system we use a bulk superconductor which is infinite
in the z and y directions and is finite in the x-direction (see
Fig. 1). This model allows us to study the effect of the
nonuniform current distribution in the superconductor (due to
transport current) on the vortex dynamics at zero and low magnetic
fields.

In our calculations we neglect the possibility of the formation of
curved vortices in the z direction and therefore our problem
becomes two-dimensional. The generalized time dependent
Ginzburg-Landau equations in our case can be written as
\begin{subequations}
\begin{eqnarray}
\frac{u}{\sqrt{1+\gamma^2|\psi|^2}} \left(\frac {\partial
}{\partial t} + \frac{\gamma^2}{2}\frac{\partial|\psi|^2}{\partial
t}
\right)\psi= \nonumber \\
=(\nabla - {\rm i} {\bf A})^2 \psi +(1-|\psi|^2)\psi,
\end{eqnarray}
\begin{equation}
 \frac{\partial {\bf A}}{\partial t} = {\rm Re} \left[\psi^*(-{\rm i}
 \nabla-{\bf A}) \psi \right] -\kappa^2 {\rm rot\,rot} {\bf A},
\end{equation}
\end{subequations}
where the parameter $\gamma=2\tau_E\Delta(T)/\hbar$ is the product
of the inelastic collision time $\tau_E$ for electron-phonon
scattering and $\Delta(T)=4k_BT_cu^{1/2}/\pi\sqrt{1-T/T_c}$ is the
value of the order parameter at temperature T which follows from
Gor'kov's derivation \cite{Gor'kov} of the Ginzburg-Landau
equations. In Eqs. 1(a,b) the physical quantities
 are measured in dimensionless units:
temperature in units of the critical temperature T$_c$, the vector
potential ${\bf A}=(A_x,A_y,0)$ and the momentum of the
superconducting condensate ${\bf p}=\nabla \phi -{\bf A}$ are
scaled in units $\Phi_0/(2\pi\xi(T))$ (where $\Phi_0$ is the
quantum of magnetic flux), the order parameter
$\psi=|\psi|e^{i\phi}$ in units of $\Delta(T)$ and the coordinates
are in units of the coherence length $\xi(T)=(8k_B(T_c-T)/\pi
\hbar D)^{-1/2}$. Time is scaled in units of the Ginzburg-Landau
relaxation time $\tau_{GL}=\pi \hbar/8k_B(T_c-T)u$, voltage (V) is
in units of $\varphi_0=\hbar/2e\tau_{GL}$ ($\sigma_n $ is the
normal-state conductivity).  In these units the magnetic field is
scaled with $H_{c2}=\Phi_0/2\pi \xi^2$ and the current density
with $j_0=\sigma_n\hbar/2e\tau_{GL}\xi$. Following Ref.
\cite{Kramer} the parameter $u$ is taken to be equal to $5.79$.

Instead of the usual gauge ${\rm div} {\bf A}=0$ we chose the
electrostatic potential equal to zero $\varphi=0$. The
semi-implicit algorithm was used \cite{Winiecki} which provides an
effective numerical solution of Eqs. 1(a,b) for the case of large
$\kappa$ values. We apply periodic boundary conditions in the y
direction $\psi(y)=\psi(y+L)$, {\bf A}(y)={\bf A}(y+L) (L is the
period - see Fig. 1) and the superconductor-vacuum boundary
conditions in the x direction $(\nabla_x-i A_x)\psi|_{x=0,W}=0$.
The transport current was introduced via the boundary condition
for the vector potential in the x - direction ${\rm rot} {\bf
A}|_z(x=0,W)=H\pm H_I$ where $H_I=2\pi I/c$ is the magnetic field
induced by the current $I$ (per unit length in the z-direction)
and $H$ is the applied magnetic field. In all our calculations we
chose $\kappa=5$ and the parameter $\gamma$ is varied from 0 to
40.

Due to the discrete nature of the vortex motion the voltage
response in our system is a time-dependent variable. We average it
over a finite time interval which is taken to be larger than the
period of the voltage variation. But this time interval can be
comparable to the switching time between different dynamic phases
and it smoothens the current-voltage characteristics at the
transition points.
\begin{figure}[hbtp]
\includegraphics[width=0.48\textwidth]{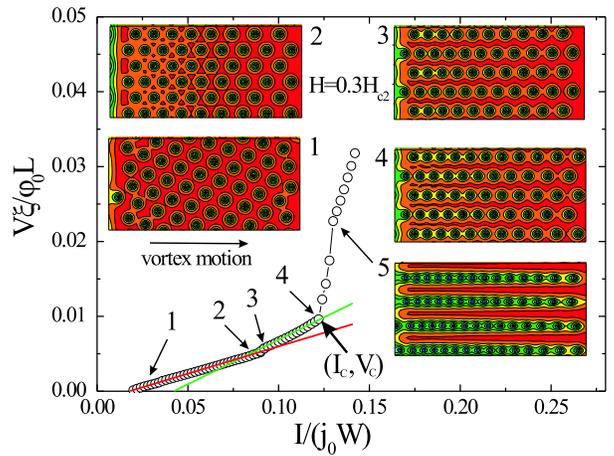}
\caption{(Color online) Current-voltage characteristics of the
superconducting slab with width $W=50 \xi$, $\gamma=10$ and $H=0.3
H_{c2}$. Current increases from zero to some finite value. In the
inset snap-shots of the order parameter at different values of the
applied current are showed.}
\end{figure}

\section{Rearrangement of the vortex lattice}

First we consider the situation when the applied magnetic field is
larger than the magnetic field due to the transport current $H \gg
H_I$. Therefore the current density distribution is almost uniform
over the width of the superconductor and it simplifies the
analysis of the obtained numerical results.  In Fig. 2 we present
the current-voltage (IV) characteristic of our system ($W=50 \xi$,
$L=25 \xi$, $\gamma=10$) under investigation at $H=0.3 H_{c2}$.
Vortex flow starts at some finite current $I_s$ (due to the
presence of the surface barrier in the system) and the vortex
structure is close to the triangular lattice. With increasing
current it transforms to a row-like structure but keeping the
triangular ordering (see point 2 in Fig. 2). Increasing the
current (arrow 2 in Fig. 2) there is a transition in the vortex
structure which is visible as a kink in the IV characteristic. The
number of vortex rows decreases (from 6 to 5 in this particular
case) and the number of vortices in the rows increases. Note that
the number of vortices in the system does not change and the kink
in the IV characteristic is connected with a faster vortex motion
in this vortex configuration. At the current indicated by the
arrow 4 there is a second transition where the number of rows
decreases further (from 5 to 4) and the distance between the
vortices in each row decreases. Simultaneously the vortex velocity
increases steeply and we have a transition to a state with a much
larger voltage.

The transitions in the vortex lattice will be explained by the
modification of the shape of the vortex core due to
non-equilibrium processes. Indeed, the motion of the vortex means
a suppression of the order parameter in front of the vortex and
recovering the order parameter behind it (see Fig. 3). If the
vortex velocity is large enough ($v \lesssim \xi/\tau_{in}$) the
number of quasiparticles in front of the vortex will be less than
the equilibrium value and larger behind the vortex due to the
finite relaxation time $\tau_{in}$ of the quasiparticle
distribution function. Effectively, we have a cooling of the
quasiparticles in front of the vortex and heating behind the
vortex (see Fig. 3). This effect is very similar to the behavior
of a superconducting weak link at voltages $V \lesssim
1/\tau_{in}$ \cite{Schmid2,Golub,Aslamazov2} when there is a
cooling at the decrease of the order parameter and a heating when
the order parameter increases in the weak link. Because the
relaxation time of the order parameter depends on the temperature
as $\tau_{|\psi|}\sim 1/(T_c-T)$ we have a long healing time of
the order parameter behind the vortex and a short time suppression
of the order parameter in front of the vortex. It leads to an
elongated shape of the vortex core with a point where $|\psi|=0$
shifted to the direction of the vortex motion. This is visible
(see Fig. 2) for vortices close to the right side of the
superconductor, where the current density and the vortex velocity
are maximal.
\begin{figure}[hbtp]
\includegraphics[width=0.48\textwidth]{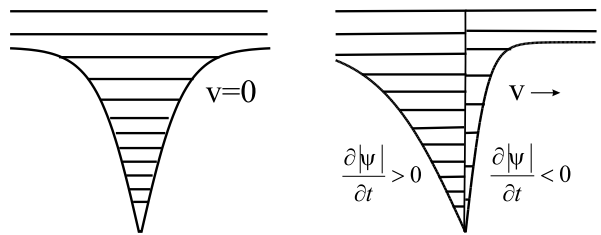}
\caption{Deformation of the vortex core due to vortex motion
(schematic). The density of the horizontal lines shows the density
of the quasiparticles. In case the diffusion length $L_{in}$ is
smaller than the coherence length $\xi(T)$ diffusion of the
quasiparticles is not strong and locally there is an effective
cooling and heating of the quasiparticles.}
\end{figure}

When the transition from 5 to 4 vortex rows in the vortex lattice
occurs the distance between the vortices suddenly decreases. If
the vortex velocity is large enough such that $v>v_c \sim
a/\tau_{|\psi|}$ ($a$ is the distance between vortices in the row)
the order parameter does not have sufficient time to recover after
every vortex passage in the row and $|\psi|$ will be strongly
suppressed along the vortex trajectory. It speeds up the vortex
motion because the time variation of $|\psi|$ depends on the value
of $|\psi|$: $\tau_{|\psi|} \sim \gamma |\psi| \tau_{GL}$
\cite{Michotte}. This is the reason for a transition from slow to
fast vortex motion (quasi-phase slip line behavior) and a steep
increase in the voltage at the point where the current is $I_c$
and the voltage is $V_c$ in Fig. 2.
\begin{figure}[hbtp]
\includegraphics[width=0.48\textwidth]{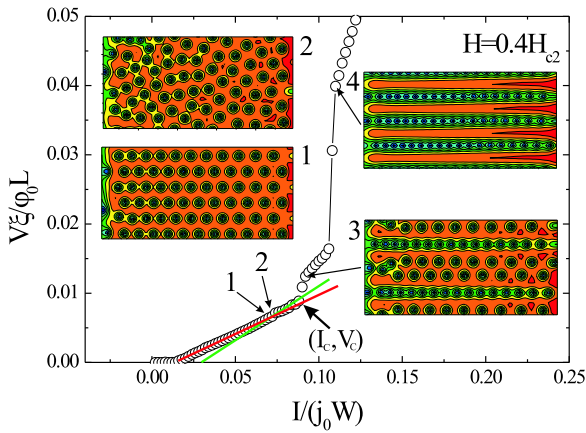}
\caption{(Color online) Current-voltage characteristics of the
superconducting slab with width $W=50 \xi$, $\gamma=10$ and $H=0.4
H_{c2}$.}
\end{figure}

At higher magnetic fields the situation is similar to the case
$H=0.3H_{c2}$ but in addition a transition to a state with vortex
rows moving with different velocities is possible (see Fig. 4).
When a fast vortex row (which we will further call a quasi-phase
slip line (PSL)) appears in the sample the superconducting current
decreases around the PSL on the scale of the decay of the electric
field $E$ (or charge imbalance) $L_E$ \cite{Tinkham}. Then
vortices in adjacent to PSL areas are forced to move with smaller
velocities because the superconducting current mainly drives them.
In the framework of the model equations 1(a,b) $L_E\simeq
\sqrt{\gamma/u} \xi > L_{in}$ \cite{Michotte} and we found indeed
that for larger values of $\gamma$ the current and magnetic field
range over which this structure may exist increases. For example,
for $\gamma=20$ the slow and fast vortex rows may coexist already
at $H=0.3 H_{c2}$ and for $\gamma=40$ they may coexist even at
$H=0$.

Note, that in contrast to the case $H=0.3 H_{c2}$ the instability
of the vortex lattice leads to quasi-chaotic vortex motion at
currents between points 2 and 3. We relate this to the usage of
periodic boundary conditions. For example in case of $H=0.3H_{c2}$
the same chaotic behavior (not shown here) disappears between
points 2 and 3 in Fig. 2 with an increase of the period of our
system by a factor of two (with a small effect on the values of
the currents where structural transitions occur). But for $H=0.4
H_{c2}$ doubling the period did not result into any effect.

We explain the influence of the boundary conditions by
incommensurability effects between the period L and $L_E$.
Actually the latter length defines the scale of the interaction
between phase slip lines. Changing the parameter $\gamma$ we
change $L_E$. For example for $\gamma=20$ and $H=0.4 H_{c2}$ we
did not observe any irregular vortex distribution for the
superconductor for the parameters corresponding to Fig. 3 even
when $L=25 \xi$.
\begin{figure}[hbtp]
\includegraphics[width=0.48\textwidth]{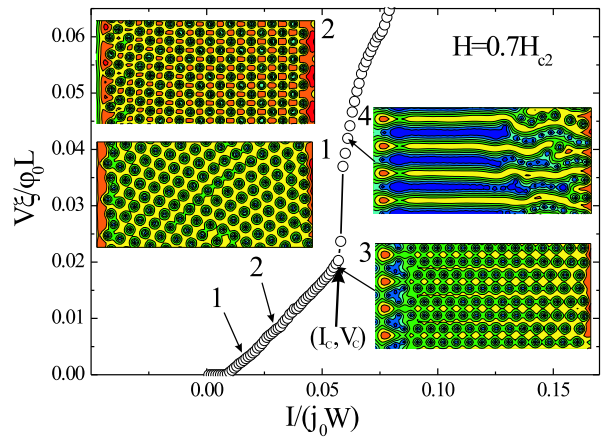}
\caption{(Color online) Current-voltage characteristics of the
superconducting slab for the parameters of Fig. 2 and $H=0.7
H_{c2}$.}
\end{figure}

At magnetic fields close to $H_{c2}$ there are also transitions in
the vortex structures, but they are masked by a large number of
possible transitions due to the increased number of vortices in
the system (see Fig. 5). The kinks in the current-voltage
characteristics become almost invisible and the jumps in the
voltage gradually decreases at the current $I_c$ where the
quasi-phase slip lines appear in the system.

\section{Vortex motion at zero magnetic field}

In Fig. 6 we present the IV characteristic of the same sample as
in Fig. 2 at zero magnetic field. At low currents we have slow
vortex motion while at large current quasi-phase slip lines
appears. We should note that we did not observe any structural
changes in the vortex lattice at low magnetic fields due to the
small number of vortices and hence the large distance between
them.
\begin{figure}[hbtp]
\includegraphics[width=0.48\textwidth]{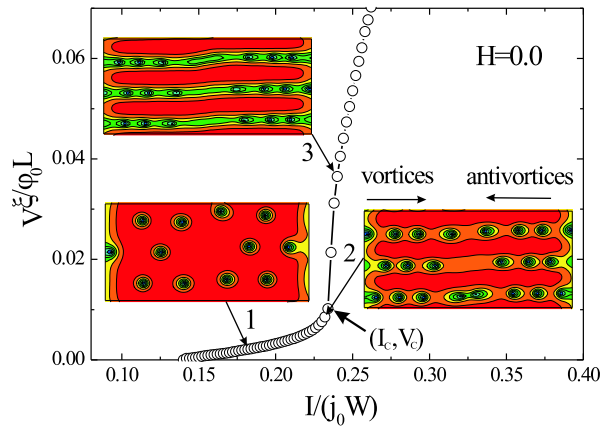}
\caption{(Color online) Current-voltage characteristics of the
superconducting slab with the same parameters as in Fig. 2, but
now for zero magnetic field.}
\end{figure}

In case of zero magnetic field the current density is strongly
nonuniform over the width of the sample (see Fig. 7). When the
current exceeds the critical value $I_s$ (current of suppression
of the surface barrier for vortex entry) the Meissner state is
destroyed, vortices and antivortices enter the sample, pass
through it and annihilate in the center. This process results in
the appearance of an additional maximum in the current density in
the center of the sample - see Fig. 7 (in agreement with
analytical calculations of Refs. \cite{Aslamazov,Blok}). For
sample parameters of Fig. 6 the quasi-phase slip behavior starts
when the current density in the center reaches the value close to
the depairing current density $j_{dep}$. We found that at this
moment the annihilation of vortex-antivortex pairs speeds up and
it provides a favourable condition for fast vortex motion across
the whole sample. However with increasing width of the sample the
transition to the fast vortex motion behavior starts at a larger
current (see Fig. 8).
\begin{figure}[hbtp]
\includegraphics[width=0.48\textwidth]{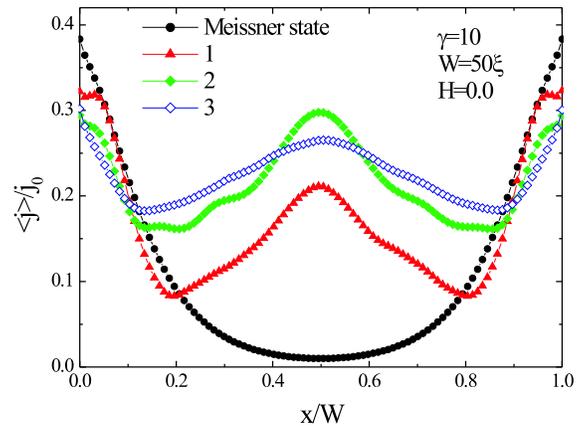}
\caption{(Color online) Distribution of the time and length
averaged current density $<j>=j_n+j_s$ over the width of the
superconducting slab at zero magnetic field and different values
of the transport current. Numbers 1-3 in the figure corresponds to
different values of the transport current in Fig. 6.}
\end{figure}
As the speed of the fleet is defined by the speed of the slowest
ship the nucleation of the quasi-phase slip line depends on the
vortex motion in the place where the current density is minimal.
When we increase the width of the sample we decrease the minimal
current density $j_{min}$ in the sample (compare Figs. 8(b) and
7). When $j_{min}$ reaches the critical value the quasi phase slip
line nucleates in the sample. This critical value is smaller the
larger $\gamma$. This result is closely connected with the
findings of Ref. \cite{Michotte} where it was shown that the phase
slip process does not exist in a quasi-1D superconductor and 2D
thin superconducting films with uniform distribution of the
current density \cite{Vodolazov} if the current density is smaller
than some critical value $j_{c1} (\gamma)$.
\begin{figure}[hbtp]
\includegraphics[width=0.48\textwidth]{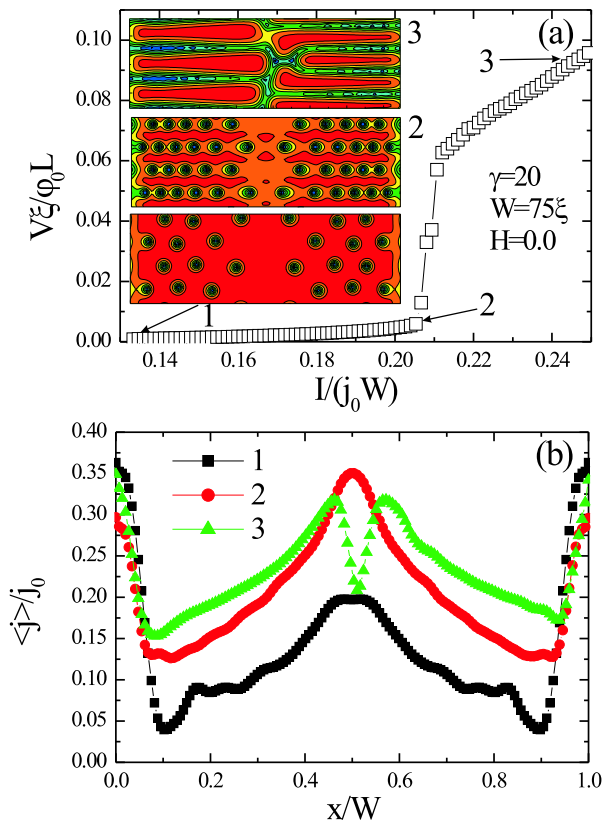}
\caption{(Color online) (a) Current-voltage characteristic of the
superconducting slab with parameters: $\gamma=20$, $W=75 \xi$ at
zero magnetic field. In the inset to Fig.(a) we present snap-shots
of $|\psi|$ at different values of the current. (b) Distribution
of the time and length averaged current density $<j>=j_n+j_s$ for
wider sample. Numbers correspond to different values of the
current at Fig. (a).}
\end{figure}

We like to stress that we did not find that the vortices and
anti-vortices can pass through each other as predicted in Ref.
\cite{Weber}. Probably the uniform current distribution used in
Ref. \cite{Weber} is essential to observe this effect.

\section{Discussion}

\subsection{Comparison with other theoretical works}

Our results strongly support the intuitive idea (published already
in Ref. \cite{Volotskaya}) about nucleation of phase slip lines at
large currents against a background of vortex flow. This idea was
further developed in the theoretical work \cite{Lempitskii} where
the current-voltage characteristics of a wide superconducting film
with both viscous vortex flow and phase slip lines was calculated.
However the author used equations that are averaged over the
inter-vortex distance and did not find the rearrangements of the
vortex structure at high vortex velocities.

In Ref. \cite{Glazman} the appearance of the wake behind the
moving vortex was theoretically predicted on the basis of an
analytical solution of the linearized equation (1a) for the
absolute value of the order parameter. Actually such a wake should
exist even in the simple time-dependent Ginzburg-Landau equation
(with $\gamma=0$) due to the finite time for the order parameter
relaxation $\tau_{|\psi|} \sim \tau_{GL}$. Indeed, when a vortex
moves the current density in front of its motion is the sum of the
current density from the vortex $j_{vort}$ and the transport
current density $j_{ext}$ and likewise behind the vortex it is the
difference $j_{vort}-j_{ext}$. The time relaxation of the order
parameter depends on the value of the current density if $j$ is
close to the depairing current density (the larger the current the
smaller is $\tau_{|\psi|}$ - see chapter 11.4 in Ref.
\cite{Tinkham}). If $j_{ext}$ is close to $j_{dep}$ the difference
between $\tau_{|\psi|}$ in front and behind the moving vortex is
substantial \cite{ours1} and the moving vortex becomes elongated
in the direction of its motion.

The change in the shape of the vortex was also found in Ref.
\cite{Andronov} on the basis of a numerical solution of the 2D
time-dependent Ginzburg-Landau equations. Such vortices were
called kinematic vortices due to their high velocity. They were
found to exist when a quasi-phase slip line nucleated in the
sample. The system resembles a Josephson vortex in a long
Josephson junction where anisotropy is connected with different
penetration depths of the screening current along and across the
Josephson junction (see chapter 6.4 in \cite{Tinkham}). In case of
a phase slip line the anisotropy is connected with a strongly
suppressed order parameter in the direction of the vortex motion
(along the quasi-phase slip line).

In both of the above works the dependence of the relaxation time
$\tau_{|\psi|}$ on the value of the order parameter was ignored.
In Ref. \cite{Andronov} the term
$\gamma^2\partial|\psi|^2/2\partial t$ on the left hand side of
Eq. 1(a) was neglected and instead of the coefficient
$u/\sqrt{1+\gamma^2|\psi|^2}$ the variable parameter $u^*$ was
used \cite{Andronov}. Actually in Ref. \cite{Glazman} the same
approach was used because in Eq. 1(a) a fixed value for the order
parameter $|\psi|$ in the term $\tau_{|\psi|}=\tau_{GL}u
\sqrt{1+\gamma^2|\psi|^2}$ was put.

But in a TDGL model with $\gamma=0$ and arbitrary value of $u$ we
did not find any steep transition from slow vortex flow to fast
vortex motion (quasi-phase slip line) at finite value of the
applied magnetic field. The reason is simple: in that model the
relaxation time of the order parameter practically does not depend
on the value of the order parameter and the mechanism of the
switching in vortex motion discussed in Sec. III does not work.
Besides we did not find any vortex structure rearrangement in the
model with $\gamma=0$ and arbitrary $u$. Probably the change of
the shape of the vortex is small in the above simplified model.

In Ref. \cite{Vodolazov} Eq. 1(a) was coupled with the equation
for the electrostatic potential and the transition from the slow
vortex flow to the quasi-phase slip lines behavior was numerically
observed in case of thin 2D superconducting films of finite length
in a perpendicular magnetic field. However the rearrangement of
the vortex lattice and the coexistence of the fast and slow vortex
motion were not found because of the small width of the samples.

\subsection{Range of validity of the obtained results}

Our results are strictly valid only when $L_{in}<\xi(T)$ while
usually in experiments $L_{in} \gg \xi(T)$. But it is obvious that
cooling and heating of the quasiparticles around the vortex core
occurs in both limits. Large $L_{in}$ provides some kind of space
averaging of these different processes (due to diffusion of the
non-equilibrium quasiparticles from the overheated region to the
overcooled one). In the framework of the LO theory the zero order
effect was calculated (when the dependence on the direction of the
vortex motion was neglected) which roughly leads to an effective
cooling of the system and a symmetrical shrinkage of the vortex
core if the distance between the vortices is much smaller than
$L_{in}$. In this limit the separation of the system into slow
moving vortices and quasi-phase slip lines is in principle
possible for $v>v_c$ when the vortex motion becomes unstable. The
origin for such a behavior is the presence of the normal component
of the current density (electric field) and its finite decay
length from the phase slip line $L_E$. The slow vortex motion
between the quasi-phase slip line may occur due to a weakening of
the superconducting component of the current near the quasi-phase
slip line. In the LO theory this effect was neglected and only
deviations of the longitudinal part of the quasiparticle
distribution function from equilibrium was taken into account
while the transverse (even in energy) part of $f(E)$ is
responsible for the appearance of the finite normal current in the
superconductor \cite{Tinkham}.

The situation is different if the distance between the vortices
exceeds $L_{in}$ and the effective averaging becomes weaker. The
anisotropy of the vortex core and the effective attraction between
vortices should be more pronounced and leads to the appearance of
vortex rows/lines (slow or fast) even at vortex velocities less
than $v_c$.

\subsection{Comparison with experiments}

The important property which follows from our calculations is the
weak dependence of the critical voltage $V=V_c$ on the applied
magnetic field (see inset in Fig. 9). We explained it by the
rearrangements of the vortex lattice when the vortex velocity
approaches $v_c$. Because in this case we do not have a triangular
vortex lattice the distance between vortices in the rows will be
smaller than $a\sim\sqrt{\Phi_0/B}$ and defined by $a \sim 1/B$
dependence. Indeed, at current $I=I_c$ the transition to a state
with 4 quasi-phase slip lines occurs (see Figs. 2,4,5) in the
magnetic field range 0.3-0.7 $H_{c2}$ while the number of the
vortices in the sample increases linearly with magnetic field.
Assuming that the transition to the fast vortex motion state
occurs when the distance between the vortices {\it in the row} is
equal to $a\sim 1/B \sim v_c \tau_{|\psi|}$ (see Sec. III) we
obtain  $v_c\sim 1/B \tau_{|\psi|}$ and the field independent
critical voltage $V_c=v_cBL$. Such a dependence was experimentally
observed in Ref. \cite{Chiaverini} for both low and high
temperature superconductors in the low magnetic field regime where
the vortex separation $a \gg L_{in}$ and the self-field of the
transport current was negligible.

From the above estimations it follows that $V_c \sim 1/\gamma$. We
see from the inset in Fig. 9 that indeed $V_c \sim 1/\gamma$ and
besides the resistivity of the superconductor at low currents
follows the dependence $\rho/\rho_n \sim 1/\gamma$ (see Fig. 9)
analytically found in \cite{Larkin3} for large magnetic fields in
the temperature interval $T_c-\hbar/k_B \tau_{in}<T<T_c$ where
equations 1(a,b) are valid.

Our estimation of the critical velocity was made in the spirit of
the paper of Doettinger et al. \cite{Doettinger3}. They supposed
that the vortex motion instability occurs when the non-equilibrium
quasiparticles induced in the vortex core does not have time to
relax to equilibrium when the next vortex arrives to the place
where they were induced $\tau_{in}=a/v_c$. Actually it means that
the order parameter did not have time to increase (because its
value depends on $f(E)$ and it cannot grow faster than
$\tau_{in}$).
\begin{figure}[hbtp]
\includegraphics[width=0.48\textwidth]{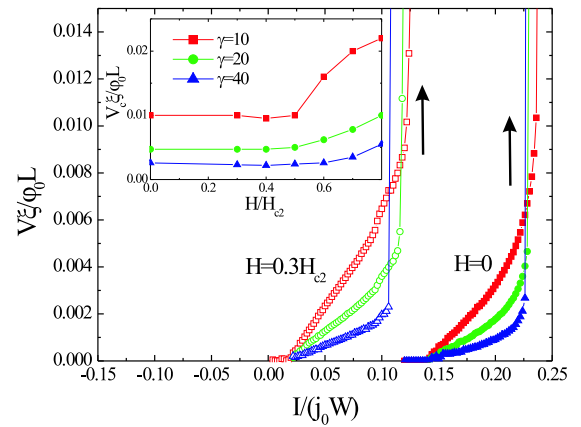}
\caption{(Color online) Current-voltage characteristics of the
superconducting slab with width $W=50 \xi$, $\kappa=5$ and
different values of the parameter $\gamma$ at two values of the
magnetic field. In the inset the dependence of the the critical
voltage on the magnetic field is shown for three values of the
parameter $\gamma=10,20,40$.}
\end{figure}

The stair-like structure of the IV characteristics (which is a
fingerprint of the nucleation of the phase slip centers or
quasi-phase slip lines) was observed both in low and high
temperature superconductors
\cite{Volotskaya,Volotskaya2,Dmitriev,Zybtsev,Sivakov} at low
magnetic fields. In wide samples (in which the strongly nonuniform
current density distribution over the width of the sample is
realized in the Meissner state) a slow vortex motion was found low
currents which is changed into the quasi-phases slip line behavior
at higher currents \cite{Volotskaya,Volotskaya2,Dmitriev,Zybtsev}.
It is interesting to note that the stair-like structure was also
experimentally observed at high magnetic fields (and temperatures
far below critical one) in high-temperature superconductors
\cite{Kunchur2,Kunchur3}. In this case one was able to observe it
only in the voltage driven regime.

In the experiments of Kunchur et. al
\cite{Kunchur,Kunchur2,Kunchur3} the quasi-phase slip lines become
visible because the penetration length of the electric field $L_E$
increases in high magnetic field due to the suppression of the
order parameter by vortices and a steep increase of $\tau_{in}$ at
low temperatures \cite{Doettinger}. At low magnetic fields and
high temperatures too many quasi-phase slip lines appear
simultaneously at $I=I_c$ and it smooths out the stair structure
of the IV characteristic.

We explain the absence of the stair-like structure in the majority
of the experiments where the LO effect was studied due to the
large length of the samples as compared to $L_E$. At the
instability point many quasi-phase slip lines should appear in
such a sample and it results into a large heating of the
electronic subsystem and the sample itself. That could be the
reason for the transition to the nearly normal state. Besides in
those samples the role of the voltage can be very pronounced and
it leads to an additional suppression of superconductivity, the
effect which is absent in the model equations 1(a,b)
\cite{Vodolazov3}. Therefore it would be interesting to perform an
experiment on a short superconducting bridge with length of about
several $L_E$ at different temperatures and magnetic fields.
Taking into account the strong dependence of $\tau_{in}$ and hence
$L_E$ on temperature it would be wort to study several samples
with different length satisfying the condition $L\sim L_E$ at
different temperatures. By variation of the magnetic field one
could observe the predicted splitting of the vortex flow into
regions with fast vortex flow (quasi-phase slip lines) and slow
vortex flow. A good candidate is NbGe which has rather low bulk
pinning even at $T \sim T_c/2$ \cite{Babic}.

\subsection{Hysteretic behavior}

The hysteresis is almost absent in zero magnetic field for a
superconducting slab with $W=50$, $\kappa=5$ and $\gamma=10$ (see
Fig. 10). Hysteresis appears when we increase the applied magnetic
field (see Fig. 10) or decrease the width of the sample. In both
cases the current density distribution becomes more uniform in the
sample and it brings hysteresis in the system.

The physical reason for this effect is as follows. Consider at
first the case of zero applied magnetic field $H=0$. In Ref.
\cite{Michotte,Vodolazov} it was found that in a superconductor
with uniform current density distribution the phase slip
center/line does not exist at current density $j_{c1}(\gamma)$
which can be smaller than $j_{dep}$ (in case of zero
fluctuations). But the superconducting state in such a system can
be stable up to $I_c=j_{dep}dW$. When a quasi-phase slip line
nucleates at $I_c$ it can exist up to a smaller current
$I_1=j_{c1}dW$ and it leads to hysteretic behavior. When we take
into account screening effects, the current density distribution
becomes nonuniform over the width of the sample. It is maximal on
the edge and minimal in the center of the sample being in the
Meissner state. At current $I=I_s<I_c$ the current density on the
edge reaches the depairing current density and the superconducting
state becomes unstable. Vortices enter the sample and if the
minimal current density is larger than $j_{c1}$ they move fast and
form a quasi-phase slip line. If the minimal current density is
smaller than $j_{c1}$ they move slowly and form quasi-phase slip
lines at a larger current when the condition $j_{min}>j_{c1}$ is
fulfilled. In the latter case hysteresis is absent because the
transition from the slow to the fast vortex motion (or vice
versus) occurs when the current density in one point (over the
width of the sample) reaches the critical value. It does not lead
to a crucial redistribution of the normal and superconducting
current density over the sample and the vortex motion is almost
non-hysteretic.

When a magnetic field is applied the transition to the fast vortex
motion occurs when the current density reaches a critical value
(which depends on the magnetic field) practically over the whole
sample. As a result the distribution of the normal and
superconducting current density changes drastically over the whole
sample and it provides the hysteretic behavior for vortex motion
in our model system.
\begin{figure}[hbtp]
\includegraphics[width=0.48\textwidth]{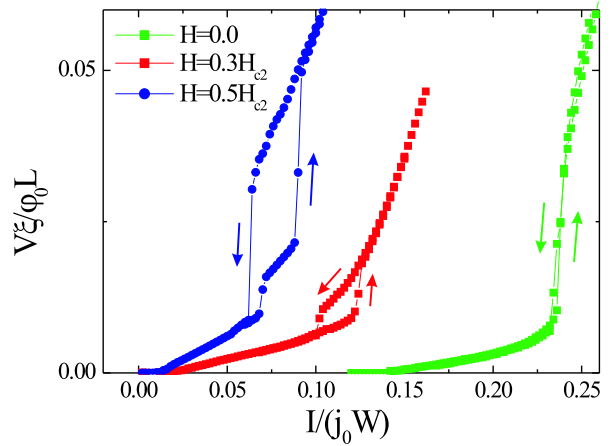}
\caption{(Color online) Hysteresis in the current-voltage
characteristics of the superconducting slab with $W=50 \xi$,
$\gamma=10$ and different magnetic fields.}
\end{figure}

In our calculations we did not take into account heating effects.
Their incorporation in the considered model will increase the
hysteresis \cite{Bezuglij} and mask all effects \cite{Vodolazov3}
discussed in our paper.

\subsection{Synchronization of vortex motion in adjacent vortex rows}

Can the motion of the vortices in two adjacent quasi-phase slip
rows/lines be synchronized? It was found in many papers (for a
review see \cite{Kadin,Tidecks}) that the dynamics of the order
parameter in one phase slip center may influence the dynamics of
the order parameter in a remote phase slip center even if the
distance between them is large. The effect is mainly connected
with the long decay length of the quasiparticle (normal) current
from the phase slip center. As a result the ac component of the
normal current affects the oscillations of the order parameter in
the other phase slip center in a way similar to a Josephson
junction under the action of an external ac current or
electromagnetic radiation. The interaction between phase slip
centers becomes more complicated if one take into account the
nucleation of the charge imbalance waves \cite{Kadin} which can
both enhance and suppress the synchronization of the the order
parameter oscillations in adjacent phase slip centers/lines
\cite{Kadin}.
\begin{figure}[hbtp]
\includegraphics[width=0.48\textwidth]{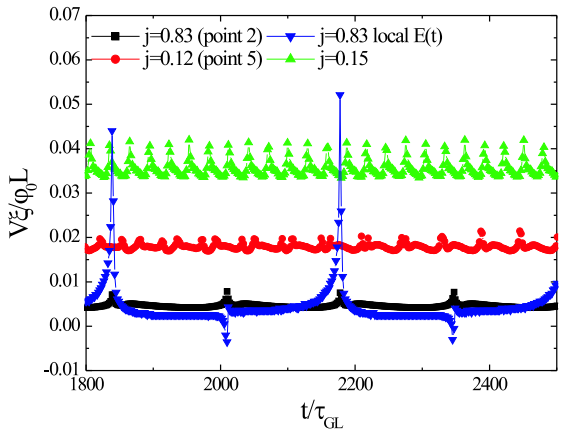}
\caption{(Color online) Time dependence of the instant voltage and
local electric field at different values of the applied current
for the sample with parameters as in Fig. 2.}
\end{figure}

In our calculations, effect due to charge imbalance waves were not
considered and quasi-phase slip lines interact only via the ac
component of the normal current. In Fig. 11 we present the time
dependence of the averaged, over the length of the sample,
electric field at the edge of the superconductor and the local
electric field in the point where the quasi-phase slip line at
large current nucleates (at $H=0.3H_{c2}$ and different currents).
At low currents, when the induced voltage and normal current are
small the motion of the vortices in adjacent rows is out of phase
because of weak interactions between vortex and strong repulsion
between vortices. Contrary, at large currents the exit of the
vortices from the superconductor in adjacent rows becomes in
phase. It means that the emitted electromagnetic radiation should
be considerably enhanced in this case in comparison with the low
current limit.

The frequency of the radiation can be tuned by an applied magnetic
field or/and by applied current. The applied magnetic field
changes the number of the vortices in the row and hence changes
the distance between vortices while transport current changes the
vortex velocity. Both factors influences the frequency of the
extracted radiation $\nu=v/a$. Taking typical values for the
critical velocity in YBCO at T=77 K and B=1 T $v_c=10^3 m/s$
\cite{Doettinger} and the inter-vortex distance $a \simeq
\sqrt{\Phi_0/B}=40 nm$ we obtain $\nu \simeq 4\cdot 10^{11} Hz$.
In the quasi-phase slip line regime the vortex velocity $v \gg
v_c$ and the frequency approaches the THz regime.

\section{Conclusions}

In the framework of the generalized time-dependent Ginzburg-Landau
equations we showed that with increasing applied current the
moving Abrikosov vortex lattice changes its structure from a
triangular one to a set of parallel vortex rows. The effect
originates from changes in the shape of the moving vortex. The
vortex core becomes elongated in the direction of vortex motion
because of different relaxation times of the order parameter in
front and behind the moving vortex. In front of the moving vortex
the order parameter may vary very fast due to a large value of the
local current density and a deficit of quasiparticles in
comparison with its equilibrium value. Contrary the number of the
quasiparticles exceeds locally their equilibrium value and the
current density is small behind the moving vortex and it increases
the relaxation time of the order parameter. This results in the
appearance of a wake behind the vortex which attracts other
vortices.

We found that the rearrangement of the vortex lattice results in
field-independent value of the critical voltage at which the
transition to the state with quasi-phase slip lines occur. This is
connected with changes of the inter-vortex distance at the
structural transitions of the vortex lattice. In a triangular
lattice the average distance between vortices varies as $a\sim
1/\sqrt{B}$ while in case of vortex rows the minimal inter-vortex
distance decreases with increasing magnetic field as $a\sim 1/B$.
It results in dependence $V_c\sim \sqrt{B}$ for triangular lattice
and $V_c\sim const$ for vortex rows.

At some magnetic field the quasi-phase slip lines can coexist with
slowly moving vortices between such lines. Besides we found that
the motion of the vortices in adjacent quasi-phase slip lines can
be synchronized at large vortex velocity $v>v_c$. Both effects are
possible due to the long decay length of the normal current near
the quasi-phase slip line. It decreases the superconducting
component of the current in the system and provides
synchronization of oscillations in the order parameter at the
quasi-phase slip lines.

Although our results are strictly valid when $\xi(T)>L_{in}$ they
qualitatively explain experiments on the instability of the vortex
flow at low magnetic fields when the distance between vortices $a
\gg L_{in} \gg \xi (T)$. Besides our results support the idea that
a similar instability of the vortex lattice should exist for
$v>v_c$ even when $a<L_{in}$.

\begin{acknowledgments}
We thank V. V. Kurin for useful discussions. This work was
supported by the Flemish Science Foundation (FWO-Vl), the Belgian
Science Policy (IAP) and the ESF-AQDJJ program. D. Y. V.
acknowledges support from INTAS Young Scientist Fellowship
(04-83-3139) and the Dynasty Foundation.
\end{acknowledgments}


\begin{references}
\bibitem{ours1} The relaxation time behind the moving vortex could be estimated
from the work of Glazman \cite{Glazman} if we replace the time
$\tau_{in}$ by $\tau_{GL}$ in his results.
\end{references}
\end{document}